\documentclass[aps,prd,groupedaddress,amssymb,twocolumn,eqsecnum,showpacs,epsfig,nofootinbib]{revtex4}

\usepackage{graphicx}
\usepackage{bm}
\usepackage{dcolumn}
\usepackage{amsmath}

\usepackage{amsmath}
\usepackage{amssymb}

\numberwithin{equation}{section}
\usepackage{epsfig}

\begin{document}
\newcommand{\newc}{\newcommand}

\newc{\be}{\begin{equation}}
\newc{\ee}{\end{equation}}
\newc{\ba}{\begin{eqnarray}}
\newc{\ea}{\end{eqnarray}}
\newc{\bea}{\begin{eqnarray*}}
\newc{\eea}{\end{eqnarray*}}
\newc{\D}{\partial}
\newc{\ie}{{\it i.e.} }
\newc{\eg}{{\it e.g.} }
\newc{\etc}{{\it etc.} }
{\newc{\etal}{{\it et al.}}
\newc{\lcdm}{$\Lambda$CDM}
\newcommand{\nn}{\nonumber}
\newc{\ra}{\rightarrow}
\newc{\lra}{\leftrightarrow}
\newc{\lsim}{\buildrel{<}\over{\sim}}
\newc{\gsim}{\buildrel{>}\over{\sim}}
\newcommand{\mincir}{\raise
-3.truept\hbox{\rlap{\hbox{$\sim$}}\raise4.truept\hbox{$<$}\ }}
\newcommand{\magcir}{\raise
-3.truept\hbox{\rlap{\hbox{$\sim$}}\raise4.truept\hbox{$>$}\ }}

%\title{Looking beyond Einstein's gravity with the evolution of linear bias}
\title{Testing General Relativity Using the Evolution of Linear Bias}

\author{S. Basilakos}\email{svasil@academyofathens.gr}
\affiliation{Academy of Athens, Research Center for Astronomy and
Applied Mathematics,
 Soranou Efesiou 4, 11527, Athens, Greece
\\ High Energy Physics Group, Dept. ECM, Univ. de Barcelona,
Av. Diagonal 647, E-08028 Barcelona, Spain}

\author{J. B. Dent}\email{jbdent@asu.edu}
\affiliation{Department of Physics and School of Earth and Space Exploration, Arizona State University, Tempe, AZ 85287-1404, USA}

\author{S. Dutta}\email{sourish.d@gmail.com}
\affiliation{Department of Physics and Astronomy, Vanderbilt University,
Nashville, TN-37235, USA}

%\affiliation{Mathematics and Natural Sciences Division, Nashville State Community College, Nashville, TN-37209, USA}

\author{L. Perivolaropoulos}\email{leandros@uoi.gr}
\affiliation{Department of Physics, University of Ioannina, Greece}

\author{M. Plionis}\email{mplionis@astro.noa.gr}
\affiliation{Institute of Astronomy \& Astrophysics, Nationals
Observatory of Athens, Thessio 11810, Athens, Greece, and
\\Instituto Nacional de Astrof\'isica, \'Optica y Electr\'onica, 72000 Puebla, Mexico}

\begin{abstract}
We investigate the cosmic evolution of the linear bias 
in the framework of a flat FLRW spacetime. We consider metric perturbations in the Newtonian 
gauge, including Hubble scale effects. Making the following 
assumptions,  (i) scale independent current epoch bias 
$b_0$,  (ii) equal accelerations between tracers and 
matter, (iii) unimportant halo merging effects 
(which is quite accurate for $z<3$), we analytically derive 
the scale dependent bias evolution. The identified scale 
dependence is only due to Hubble scale evolution GR effects, while 
other scale dependence contributions are ignored. We find that up 
to galaxy cluster scales the fluctuations of the metric do not 
introduce a significant scale dependence in the linear bias. 
Our bias evolution model is then used to derive a connection 
between the matter growth index $\gamma$ and the observable 
value of the tracer power spectrum normalization $\sigma_8(z)$.
We show how this connection can be used as an observational test 
of General Relativity on extragalactic scales.

%In this article we investigate the cosmic evolution of the linear bias
%in the framework of a flat FLRW spacetime, by taking into account
%metric perturbations using the Newtonian gauge approach, and
%deriving an analytical solution by solving a second order differential equation.
%We find that up to galaxy cluster scales
%the fluctuations of the metric do not introduce a significant scale
%dependence in the linear bias.
%Armed with our bias evolution model it is straightforward
%to obtain theoretically the cosmic structure growth history 
%which can then be compared with observational data of extragalactic
%objects to test the validity of general relativity on cosmological
%scales.
%An excellent probe for this kind of comparison has been shown to be the 
%redshift evolution of the combination of two quantities: the 
%growth rate of structure and the rms fluctuations of the linear density field.  

\end{abstract}
\maketitle

\section{Introduction}
The distribution of matter on
large scales, based on different extragalactic objects,
can provide important constraints on models of cosmic structure formation.
However, a serious problem that hampers such a straightforward
approach is our limited knowledge of how luminous matter traces the
underlying mass distribution. In particular,
the concept of the so called {\it biasing} between different classes
of extragalactic objects and the background matter distribution was
put forward by Kaiser \cite{Kai84} and Bardeen et al. \cite{Bar86}
in order to explain the higher amplitude of the 2-point correlation function
of clusters of galaxies with respect to that of galaxies themselves.

Such a biasing is statistical
in nature; with galaxies and clusters being identified as high peaks
of an underlying, initially Gaussian, random density field, while it
is assumed to be linear and scale-independent\footnote{
However, in the non-linear 
scales of clustering, ie., mostly below 5 $h^{-1}$ 
Mpc, a scale dependence of the bias has been observed 
(see for example \cite{Som01}).}.
Formally, the linear bias factor, $b$, is defined as the ratio
of the extragalactic mass tracer
fluctuations, $\delta_{\rm tr}$, to those
of the underlying mass, $\delta_{m}$:
\be
\label{eq:1}
\delta_{\rm tr} = b \delta_{m} \;\;.
\ee
Since the two-point correlation function in a continuous density field
is defined as $\xi(r)=\langle \delta({\bf x}) \delta({\bf x}+{\bf
  r})\rangle$, one can write the bias factor as the square root of the ratio of
the two-point correlation function of the tracers to the underlying mass:
\be\label{bias1}		
b=\left(\frac{\xi_{\rm tr}}{\xi_{m}}\right)^{1/2}\;.
\ee
in which case one considers the large-scale correlation function, ie.,
%{\bf scales $\magcir 5 h^{-1}$ Mpc)}.
%scales $\magcir 1 h^{-1}$ Mpc).
scales corresponding roughly to the so-called halo-halo term of the
dark matter (hereafter DM) halo correlation function (see for example 
\cite{Hamana}).
Furthermore, since the variance of a density
field, smoothed at some scale $R$, is the correlation function at zero lag
($\sigma_R^2=\xi_R(0)=\langle \delta_R^2({\bf x})\rangle$) 
one can also write the
linear bias factor as the ratio of the variances of the tracer and
underlying mass density fields, smoothed at some linear scale,
traditionally taken to be $8 \; h^{-1}$ Mpc (at which scale the
variance is of order unity):
\be\label{bias2}		
b=\frac{\sigma_{8,\rm tr}}{\sigma_{8, m}}\;.
\ee
%since $\sigma^2_{8}=\xi(0)=\langle \delta^2({\bf x})\rangle$.
%For the benefit of the reader 
We refer the reader %would like to point here that 
to a thorough discussion of the methods to estimate 
the tracer correlation function 
and also of the corresponding determination of the 
tracer bias values, at different redshifts, 
that appears in Papageorgiou et al. 
(\cite{papag12}; and references therein).

The bias factor may have many dependencies; even assuming that it is scale
independent, it necessarily depends on the
type of the mass tracer as well as on the epoch $z$,
since the fluctuations
evolve with time as gravity draws together galaxies and mass.
It is evident, therefore, that the bias
factor should also depend on the different dark energy models
(hereafter DE), including those of modified gravity \cite{BasPl11}.
It is the redshift evolution of bias, $b(z)$, which
is very important in order to relate observations with models of
structure formation and it has been shown to be
a monotonically increasing function of redshift \cite{press}-\cite{Bas08}.

In the literature there are two basic
families of analytic bias evolution models.
The first, called the  {\em galaxy merging} bias model, 
%utilizes the halo mass function and 
is based on the Press-Schechter \cite{press} formalism,
on the peak-background split \cite{Bar86} and
on the spherical collapse model \cite{Cole},
and reproduces relatively
well the results of numerical simulations, although differences have
been found especially at the high and low DM halo mass range.
These differences have lead to modifications
of the original model to include the effects of ellipsoidal collapse
\cite{she01}; to new values of the bias
model parameters \cite{Jing98}; to new forms
of the bias model fitting function \cite{Sel04}
or even to a non-Markovian
extension of the excursion set theory \cite{Ma11}.

The second family of bias evolution models assumes a continuous
mass-tracer fluctuation field, proportional to that of the underlying
mass, and the tracers act as ``test particles''. In this context, the
hydrodynamic equations of motion and linear perturbation theory are
used. 
An original suggestion, named {\em galaxy  conserving} bias model used
the continuity equation and
the assumption that tracers and underlying mass share the same velocity
field \cite{Nus94,Fry96,Teg98,Hui07}, while 
the bias evolution is provided by the solution of a 1st order
differential equation as:
$b(z)=1+(b_{0}-1)/D(z)$,
with $b_{0}$ the bias factor at the present time and $D(z)$ the
growing mode of density perturbations.
However, this bias model suffers from
two fundamental problems: {\it the unbiased problem} ie., the fact that
an unbiased set of
tracers at the current epoch remains always unbiased in the past,
and {\it the low redshift problem} ie., the fact that this
model represents correctly the bias evolution only at relatively low
redshifts $z\mincir 0.5$ \cite{Bagla98}. Note that \cite{Simon05}
has
extended this model to also include an evolving mass tracer population
in a $\Lambda$CDM cosmology.

An attempt to derive a bias evolution model free of the above
mentioned problems, utilized all three hydrodynamical equations of
motion, linear perturbation theory and the fact that
mass-tracers and underlying mass share the same
gravity field, but not necessarily the same velocity field and that
the linear bias is scale independent.
This resulted in a second order
differential equation in $b$, the solution of which provided  
the evolution of the linear and scale-independent bias
(see \cite{BasIni,Bas01} and \cite{Bas08}). 
We would like to stress here that the provided solutions apply to
cosmological models, within the framework of general relativity.

In the context of a scale-independent bias factor, an extension of the
previous model, valid for all DE and modified gravity cosmologies, 
was recently proposed by Basilakos, Plionis \& Pouri \cite{BasPl11}.
This extension provides a tool, using the evolution of bias,
to put constraints on those DE models which adhere to general relativity,
as well as to investigate whether the DE reflects on the nature of gravity
(``geometrical dark energy'').

Overall, the scope of the present article is (a) to extend
the original Basilakos et al. \cite{BasPl11}
bias solution, by taking into account possible contributions from
the metric fluctuations,
and (b) to propose new tools that can be used
in order to test the validity of general relativity
on cosmological scales. 
%Furthermore,
%we would like to investigate how the main assumptions
%of the linear perturbation theory entered in the current methodology
%affect the evolution of bias. 

The structure of our paper is as follows.
The basic theoretical elements of the problem
are presented in section II, where we introduce [for a spatially flat
Friedmann-Lema\^\i tre-Robertson-Walker (FLRW) geometry] the basic
cosmological equations.
The issue related with the
linear bias is discussed in section III. In this section
we also present the general bias solution
in the framework, by taking into account
metric perturbations 
for the Newtonian gauge.
Finally, the main conclusions are
summarized in section V.

\section{Scale Dependent matter and tracer Density Perturbations}
Let us derive the basic equations that govern the evolution of
the mass density contrast as well as of the extragalactic
tracers, modeled here as a presurreless fluid ($p_{m}=p_{\rm tr}=0$). Note that
the perturbed FLRW spacetime in the Newtonian gauge is given by
\be
\label{space}
ds^{2}=-(1+2\Phi)dt^{2}+(1-2\Phi)a^{2}(t)d{\bf x}^{2}
\ee
where $\Phi$ is the Newtonian potential, $a(t)$ is the scale factor
(normalized to unity at the present time) and $d{\bf x}^{2}$ is
the flat spatial metric. In the current paper we assume
a slowly varying gravitational potential $\Phi$.
Note that considering the unperturbed spacetime one can easily derive the
background equations

\begin{equation}
\label{fried}
H^{2}= \frac{8\pi G}{3}(\rho_{m}+\rho_{de})\;,
\end{equation}

\begin{equation}
\label{fried1}
\dot{\rho}+3H(\rho+p_{de})=0 \;.
\end{equation}
In the above set of differential equations, an over-dot denotes
derivative with respect to time, $\rho_{m}$ and $\rho_{de}$, are
the matter and dark energy densities with
$\rho=\rho_{m}+\rho_{de}$, $H ={\dot {a}}/a$ is
the Hubble parameter, whereas $p_{de}=w_{de}\rho_{de}$, corresponds to the
pressure assuming non-clustering dark energy.
Note that for $w_{de}(z)=-1$ we recover the concordance $\Lambda$CDM model.
%It is well known that for small scales (much smaller than the horizon)
%the dark energy component
%is expected to be smooth and thus it is
%fair to consider perturbations only on the matter component of the
%cosmic fluid.

\subsection{Matter density perturbations}
In this section, we discuss
the basic equation which governs the evolution of the matter
perturbations up to horizon scales and within the framework of any
DE model. Following the notations of Dent et al. \cite{Dent:2009wi}
the perturbed (anisotropic stress-free) equations in the Newtonian gauge
take the form
\ba
\label{grper1}\ddot{\Phi}&=&-4H\dot{\Phi}+8\pi G \rho_{de} w_{de}\Phi\\
\label{grper2}\dot{\delta}_{m}&=&3\dot{\Phi}+\frac{k^2}{a^2}v_{f,m}\\
\label{grper3}\dot{v}_{f,m}&=&-\Phi
\ea
with constraint equations
\ba
\label{grcons1}3H(H\Phi+\dot{\Phi})+
\frac{k^2}{a^2}\Phi&=&-4\pi G\rho_{m}\delta_{m}\\
\label{grcons2}(H\Phi+\dot{\Phi})&=&-4\pi G\rho_m v_{f,m}
\ea
where %$\Phi$ is the Newtonian potential and 
$v_{f,m}\equiv-v_{m} a$
($v_{m}$ is the velocity potential for matter).
%peculiar velocity with respect to the general expansion).
In this context, the combination of the
relativistic equations (\ref{grper1})-(\ref{grcons2})
obtains the basic differential equation
\be
\label{odedelta}
\ddot{\delta}_{m}+ 2H\dot{\delta}_{m}+\frac{k^{2}}{a^{2}}\Phi=0\;.
\ee
A solution of the above equation provides the evolution of the matter
fluctuations in the linear regime.

On the other hand, the linear matter overdensity
$\delta_{m}\equiv \delta\rho_m/\rho_m$
is written as a function
of the gravitational
potential $\Phi$ and the background variables as
follows \cite{mabertschinger}:
\be
-4\pi G\rho_{m}\delta_{m} = \frac{k^2}{a^2}\Phi +3H^2\Phi +3H\dot{\Phi}
\label{drhom}
\ee
where $G$ denoting Newton's
gravitational constant.
In the sub-Hubble (small scale) approximation
($\frac{k^2}{a^2} \gg H^2$) equation (\ref{drhom}) takes the form
\be
-4\pi G\rho_{m}\delta_{m} = \frac{k^2}{a^2}\Phi
\label{drhomss}
\ee
which is the usual Poisson equation.
In this context, inserting the Poisson equation
(\ref{drhomss}) into eq.(\ref{odedelta}) we derive
the well known scale independent equation
\be
\label{odedelta11}
\ddot{\delta}_{m}+ 2H\dot{\delta}_{m}-4 \pi G \rho_{m} \delta_{m} =0 \;.
\ee
Now, for any type of dark energy
an efficient parametrization
of the matter perturbations
is based on the growth rate of clustering
\cite{Peeb93}
\be
\label{fzz221}
f_{0}(a)=\frac{d{\rm ln}\delta_{m}(a)}{d{\rm ln}a}=\Omega^{\gamma}_{m}(a)
\ee
where $\gamma$ is the so called growth index
(see Refs. \cite{Wang98,Linjen03,Lue04,Linder2007,Nes08})
and $\Omega_{m}(a)=\Omega_{m}a^{-3}/E^{2}(a)$.\footnote{$\Omega_{m}$ is the
density parameter at the present time and $E(a)=H(a)/H_{0}$ is the
normalized Hubble function. For the usual $\Lambda$ cosmology we have
$E(a)=(\Omega_{m} a^{-3}+1-\Omega_{m})^{1/2}$.}
Integrating eq.(\ref{fzz221}) we obtain an approximate solution
of eq.(\ref{odedelta11})
which is valid for any type of dark energy\footnote{Since the pure
matter universe (Einstein de-Sitter) has the solution of
$\delta_{ES,m}=a$, we normalize our DE models 
to get $\delta_{m}\simeq a$ at large redshifts, which should hold due to the dominance of
the non-relativistic matter component.}:
\begin{equation}
\label{Dz221}
\delta_{m}(a)=a {\rm exp} \left[\int_{a_{i}}^{a} \frac{dx}{x}\;
\left(\Omega_{m}^{\gamma}(x)-1\right) \right]
\end{equation}
where $a_{i}$ is the scale factor of the universe 
at which the matter component dominates the cosmic fluid
(here we use $a_{i} \simeq 10^{-2}$).
Following standard lines we have
$\delta_{m}(a) \propto D(a)$, where $D(a)$ is the linear growing mode,
usually scaled to unity at the present epoch $D(a)=\delta_{m}(a)/\delta_{m}(1)$.
It is interesting to mention that measuring
the growth index could provide an efficient
way to discriminate between modified gravity models and DE models
which adhere to general relativity. Indeed it was theoretically
shown that for DE models
inside general relativity the growth index $\gamma$ is well fit by
$\gamma_{\rm GR}\approx 6/11$
(see \cite{Linder2007,Nes08}).

For the benefit of the reader we point here
that for the traditional $\Lambda$ cosmology it has been found, by some of us
\cite{Dent:2008ia,Dent:2009wi}, that the linear matter fluctuation field
starts to become scale-dependent, due to metric perturbations in Newtonian gauge, on scales larger
than about $\sim 50-100h^{-1}$Mpc ($k<0.01-0.02 h$Mpc$^{-1}$).
Therefore, for large scales we have to use the generalized Poisson
equation (\ref{drhom}) which is valid up to horizon scales. 
%Notice that
%for dimensional purposes (for more details see \cite{Dent:2009wi}) we consider
%that the quantity $3H\dot{\Phi}$, which
%appeared in Eq. (\ref{drhom})
%obeys $3H\dot{\Phi}\simeq 3H^{2}\Phi$. 
Notice that on dimensional grounds we may approximate the quantity
$3H\dot{\Phi}$ in Eq. (\ref{drhom}) as $3H\dot{\Phi}\simeq 3H^{2}\Phi$
(see also \cite{Dent:2009wi}). This is justified on the basis of Eq.
\ref{grper1} since [given also that $4\pi G \rho_{de} =O(H^2)$] the
only timescale that determines the evolution of $\Phi$ is the Hubble
scale $H$. It is therefore a good approximation to assume that $\dot \Phi
\simeq H \Phi$.
Using the latter condition
and inserting Eq.(\ref{drhom}) into Eq.(\ref{odedelta}) one can easily find
that
\be
\label{odedelta3}
\ddot{\delta}_{m}+ 2H\dot{\delta}_{m}-4 \pi G_{eff} \rho_{m} \delta_{m}=0
\ee
where
\be
G_{eff}(a,k)= \frac{G}{1+\xi_{k}(a,k)}
\ee
and
\be
\xi_{k}(a,k)= \frac{3a^{2}H^{2}(a)}{c^{2}k^{2}} \;.
\ee

For many DE models, it is convenient to study the
growth evolution in terms of the expansion scale $a$ rather than $t$.
If we change the variables from $t$ to $a$ ($\frac{d}{dt}=aH\frac{d}{da}$)
then the time evolution
of the mass density contrast (see Eq.~(\ref{odedelta3})) takes the
following form
\be
\label{odedelta1}
\frac{d^{2}\delta_{m}}{da^{2}}+A(a)\frac{d\delta_{m}}{da}-
B(a,k)\delta_{m}=0
\ee
where
\be
\label{afun1}
A(a)=\frac{d{\rm ln}E}{da}+\frac{3}{a}
\ee
and
\be
\label{afun2}
B(a,k)=\frac{3\Omega_{m}}{2a^{5}E^{2}(a)}\;\left[ (1+\xi_{k}(a,k)\right]^{-1} \;.
\ee

We would like to end this section with a discussion on the evolution
of the scale dependent growth rate of clustering.
Obviously, it becomes important
to construct a scale-dependent
parametrization that is analogous to Eq.(\ref{fzz221}) and solves
(approximately) Eq.(\ref{odedelta1})
for all scales $k$. In order to construct such a
parametrization
we focus on the matter dominated era when
most of the growth occurs and express $\xi_{k}(a,k)$ as \cite{Dent:2009wi}
\be
\xi_{k}(a,k)=\frac{3 H_0^2 \Omega_{m}}{a c^{2}k^2} \;.
\label{xidef2}
\ee
In this context, Dent et al. \cite{Dent:2009wi}
proposed that the scale-dependent
growth rate $f(a,k)$ may be expressed in
terms of the scale-independent growth rate $f_{0}(a)$ in the form:
\be
\label{fzz222}
f(a,k)=\frac{d{\rm ln}\delta_{m}(a,k)}{d{\rm ln}a}=
\frac{f_{0}(a)}{1+\xi_{k}(a,k)}=
\frac{\Omega^{\gamma}_{m}(a)}{1+\xi_{k}(a,k)} \;.
\ee
Thus from Eq.(\ref{fzz222}) we simply get
\begin{equation}
\label{Dz221}
\delta_{m}(a,k)=a {\rm exp}\left[\int_{a_{i}}^{a} \frac{dx}{x}
\left(\frac{\Omega_{m}^{\gamma}(x)-1}{1+\xi_{k}(x,k)}\right) \right] \;.
\end{equation}
Due to $\delta_{m}(a,k)\propto D(a,k)$ the normalized
growth factor becomes
\be
\label{Dzz}
D(a,k)=\frac{\delta_{m}(a,k)}{\delta_{m}(1,k)}=
\frac{\delta_{m}(z,k)}{\delta_{m}(0,k)}
\ee
where $a=(1+z)^{-1}$.

\subsection{Tracer density perturbations}
Now we use the same formalism as before but for the tracers.
We would like to spell out what are the basic assumptions here (see
also \cite{Padmanabhan1993}).
First of all we consider that the mass tracer population
is conserved with time i.e. that the effects of hydrodynamics
(merging, feedback mechanisms etc) do not
significantly alter the population mean. However, the effects of
merging have been phenomenologically modeled and it has been found
(using N-body simulations)
that they are important only for $z\magcir 2.5-3$ \cite{Bas08} as far as 
the bias factor is concerned.
In what follows we also treat tracers with $p_{\rm tr}=0$ which
implies that galaxies (or clusters of galaxies) are collisionless.
Thus the tracer density evolves as
\be
\label{evoltracers}
\dot{\rho}_{\rm tr}+3H\rho_{\rm tr}=0 \;.
\ee
Due to the same gravity field the
corresponding equations
(\ref{grper1}) and (\ref{grcons1}) are also valid here.
On the other hand we have
\ba
\label{grper22}\dot{\delta}_{\rm
  tr}&=&3\dot{\Phi}+\frac{k^2}{a^2}v_{f,{\rm tr}}\\
\label{grper33}\dot{v}_{f,{\rm tr}}&=&-\Phi \\
\label{grcons22} (H\Phi+\dot{\Phi})&=&-4\pi G\rho_{\rm tr} v_{f,{\rm tr}}
\ea
where $v_{f,{\rm tr}}\equiv-v_{\rm tr} a$ and
$v_{\rm tr}$ is the velocity potential of the tracers (in general
different with that of matter
$v_{\rm tr} \ne v_{m}$), $\rho_{\rm tr}$ is the tracer density, $\rho_{m}$ is
the mass density and $\Phi$ is the gravitational potential.

Since the tracers and the underlying matter share the same gravitational
field, this implies that the generalized
Poisson equation (\ref{drhom}) remains practically the same.
In other words we could have different velocity fields
($v_{\rm tr} \ne v_{m}$) but the corresponding accelerations
($\dot{v}_{\rm tr} = \dot{v}$) are the same.
Again, by taking the approximation $3H\dot{\Phi}\simeq 3H^{2}\Phi$
and using Eqs. (\ref{evoltracers}),
(\ref{grper1}), (\ref{grcons1})
and Eqs.(\ref{grper22})-(\ref{grcons22}), we can obtain after some algebra
the time evolution
equation for the tracer fluctuation field
\begin{equation}
\label{eq:12}
\ddot{\delta}_{\rm tr}+2H\dot{\delta}_{\rm tr}-4\pi G_{eff} \rho_{m} \delta_{m}=0
\end{equation}
or
\be
\label{odedelta12}
\frac{d^{2}\delta_{\rm tr}}{da^{2}}+A(a)\frac{d\delta_{\rm tr}}{da}-
B(a,k)\delta_{m}=0 \;.
\ee
\subsection{Gauge Dependence}
The Newtonian gauge used in the above calculations is physically
interesting because it corresponds to a time slicing of isotropic
expansion. However, the matter density perturbation $\delta_m(t,k)$ is
a gauge dependent quantity and therefore it is important to clarify
how our results change in alternative gauges, and what their
connection is with observable gauge invariant quantities.

Scalar metric perturbations around a spatially flat background can be
written in the following general form
\cite{mukhanov}
\begin{widetext}
\begin{equation}
  ds^2=a^2\{(1+2\Phi)d\tau^2-2B_{\mid i}dx^id\tau-[\delta_{ij}
-2(\Psi\delta_{ij}-E_{\mid_{ij}})]dx^idx^j~\},
\end{equation}
\end{widetext}
where $a$ and $\tau$ are the conformal cosmic expansion scale factor
and the conformal cosmic time; ``$_\mid$'' denotes the background
three-dimensional covariant derivative. The corresponding perturbed
energy-momentum tensor $T^{\mu}_{\nu}$ has the form
\begin{eqnarray}
  \nonumber T^0_{~0}&=&\rho_m(1+\delta_m)~,\\
  \nonumber T_0^{~i}&=&\rho_m U_{\mid i}~,\\
  \nonumber T^0_{~i}&=&-\rho_m (U-B)_{\mid i}~,\\
  T^i_{~j}&=&-\rho_m
        \Sigma_{\mid ij}~,
\end{eqnarray}
where $\rho_m$ is the unperturbed pressureless matter density; $U$ and
$\Sigma$ determine velocity perturbation and anisotropic shear
perturbation.

Gauge choices simplify the above expressions by setting various
quantities to 0. For example we have $\Psi=\Phi=B=0$
in the Synchronous Gauge ($SG$),  $B=E=0$ in the Newtonian Gauge
($NG$) and $U=B=0$ in the Comoving Time-orthogonal Gauge ($CTG$).

In the special case of the synchronous gauge, which corresponds to a
time slicing obtained by the matter local rest frame everywhere in
space (the free falling observer frame), the line element of the
perturbed spacetime is given by
\begin{eqnarray}
ds^2 = a^2(\tau)[-d\tau^2 + (\delta_{ij} + h_{ij})dx^idx^j]
\end{eqnarray}
It is straightforward to derive the growth equation for $\delta_m
\equiv \delta\rho_m/\rho_m$ in a matter dominated universe in the
synchronous gauge to obtain \cite{mabertschinger,Dent:2008ia} 
\be
\label{sgrowth}
\ddot{\delta}^{SG}_m + 2H\dot{\delta}^{SG}_m -4\pi G \rho_m 
\delta^{SG}_m = 0
\ee
This growth equation is exact in the synchronous gauge in the case of
matter domination and involves no scale dependence as in the case of
equation  (\ref{odedelta3}) of the Newtonian gauge. This scale
independence is an artifact of the particular time slicing of the
synchronous gauge which is a good approximation on small scales but is
unable to capture the horizon scale effects modifying the growth
function on large scales.

Nevertheless equations (\ref{odedelta3}) and (\ref{sgrowth}) clearly
agree on small scales where $\xi_{k}\rightarrow 0$. Therefore, for larger
scales ($k<0.01 h Mpc$) the question that arises is the following:
{\it What is the proper gauge to use when comparing with
  observations?} 

This question has been addressed in Ref. \cite{Yoo:2009au} where a
gauge invariant observable replacement was obtained for
$\delta_m$. This observable ${\delta}^{obs}_m(t,k)$ involves the matter
density perturbation $\delta_m(t,k)$ corrected for redshift
distortions due to peculiar velocities and gravitational potential. It
also includes volume and position corrections. The final expression
however is complicated and makes the theoretical predictions based on
it not easy to implement and manipulate. However, in
Ref. \cite{Chisari:2011iq} it was pointed out that the Newtonian gauge
matter perturbation ${\delta}^{NG}_m$ is a good approximation to the
observable gauge invariant perturbation ${\delta}^{obs}_m(t,k)$ even
on very large scales (comparable to the Hubble scale). This result
will also be justified in the remaining part of this section.

The cosmological perturbations evolution is well described by the
gauge-invariant (GI) approach,
pioneered by Bardeen\cite{bardeen2}. This approach may be used to
identify physical observables as gauge invariant quantities (e.g.,
Refs.\cite{kodama1,mukhanov}). A gauge-invariant matter density
perturbation may be constructed as \cite{bardeen2,kodama1}
\begin{equation}\label{epsilonm}
  \delta_m^{GIS}\equiv \delta_m +3
        \frac{\dot{a}}{a}(U-B)~,
\end{equation}
$\delta_m^{GIS}$ coincides with the density perturbation $\delta_m^{SG}$ in the
Synchronous Gauge for the pressureless matter
system. Thus $\delta_m^{GIS}$ corresponds to the density perturbation
relative to
the observers everywhere comoving with the matter. These {\it free
  falling} observers do not experience the
isotropic expanding background of the universe because the peculiar
velocity of matter is distinct from the Hubble flow. Thus
$\delta_m^{SGI}$ has physical significance only for perturbations on
scales small compared to the Hubble scale.

In addition to $\delta_m^{GIS}$, it is straightforward to construct
alternative gauge invariant quantities related to the matter
overdensity as evaluated in different gauges. Such a gauge-invariant
variable closely related to the matter overdensity in the Newtonian
gauge is of the form
\cite{bardeen2,kodama1,mukhanov},
\begin{equation}\label{gidp}
  \delta_m^{GIN}\equiv
  \delta_m+\frac{\dot \rho}{\rho}(B-\dot{E})=
\delta_m-3\frac{\dot{a}}{a}(B-\dot{E})~.
\end{equation}
and has important advantages over $\delta_m^{GIS}$ discussed in what follows.
$\delta_m^{GIN}$ coincides with the density perturbation
$\delta_m^{NG}$ in the Newtonian Gauge
($NG$), in which $B=E=0$.

It is also straightforward to construct
two gauge-invariant scalar potentials $\phi$ and
$\psi$,
which reduce to the  gravitational potential
in the Newtonian limit:
\cite{mukhanov}:
\begin{eqnarray}\label{potential}
\nonumber && \phi \equiv \Phi-\frac{\dot{a}}{a}(B-\dot{E})~,\\
      && \psi \equiv \Psi+\frac{1}{a}\frac{d}{d\tau}[(B-\dot{E})a]~.
\end{eqnarray}

The gauge invariant gravitational potential $\phi$
obeys the Poisson equation\cite{bardeen2,kodama1} sourced by
$\delta_m^{GIS}$ with no appearance of the Hubble scale:
\begin{equation}\label{poisson}
\bigtriangledown^2\phi=-k^2\phi
=4\pi G\rho a^2\delta_m^{GIS}.
\end{equation}
where $k$ is the (comoving) wavenumber of the Fourier mode. The Poisson
equation is valid only for scales small
compared to the Hubble radius $1/H$ while on scales larger than the
Hubble scale the growth of matter density perturbations is
frozen. Hence, $\delta_m^{GIS}$ can not be regarded as the observable
matter density perturbation on scales comparable to
the Hubble scale.  Therefore,
the observable density perturbation on both the small-scale and
the large-scale modes can not be described {\em directly} by
$\delta_m^{GIS}$ even though it is a gauge-invariant quantity.

The other gauge invariant perturbation $\delta_m^{GIN}$ has some
important attractive features with respect to observability, not
shared by $\delta_m^{GIS}$. These are summarized as
follows: \begin{itemize} \item It reduces to the Newtonian gauge
  perturbation $\delta_m^{NG}$, {\it i.e.} it corresponds to a frame
  which respects the isotropic expansion of the universe and is
  therefore more appropriate for description of large scale
  perturbations. This reduction also simplifies the calculation of
  this perturbation. \item It drives a scale dependent modification of
  the Poisson equation for the gauge invariant potential
  $\phi$. Indeed, the time-time part of the
linearized Einstein equation gives \cite{mukhanov,mabertschinger}
\begin{equation}\label{zero}
\bigtriangledown^2\phi-3\frac{\dot{a}}{a}(\frac{\dot{a}}{a}\psi+\dot{\phi})
%=-k^2\phi-3\frac{\dot{a}}{a}(\frac{\dot{a}}{a}\psi+\dot{\phi})
=4\pi G\rho a^2\delta_m^{GIN}.
\end{equation}
Thus, the anticipated scale dependence on scales comparable to the
Hubble scale is picked up by the perturbation $\delta_m^{GIN}$. 
\item It is gauge invariant as anticipated for any observable
quantity. 
\end{itemize}

Thus, the gauge invariant $\delta_m^{GIN}$ and the Newtonian gauge
variable $\delta_m^{NG}$ to  which it reduces, constitute an
attractive choice for making theoretical calculations to obtain the
gravitational potential and the matter density perturbation that can
be directly compared with observations on large scales. However, these
theoretically obtained quantities need to also be corrected for bias,
redshift distortions (due to gravitational potential and peculiar
velocities), lensing magnification and volume
distortion\cite{Yoo:2009au}.

\section{The evolution of bias}
Clearly, due to the metric perturbations, 
equations (\ref{odedelta3}) and (\ref{eq:12}) involve a
scale $k$ dependence of bias in contrast to the small scale approximate
equation (\ref{odedelta11}) which is scale-invariant. Therefore,
due to Eq.(\ref{eq:1}) one
would expect that the bias factor must inherit a similar
dependence to that of the density fluctuations
namely $b=b(z,k)$.
Within this framework, we
can distinguish three possible bias evolution cases:

\underline{\it Case 1: Tracers and Mass share same velocity field:}
Here we use the assumption of
Tegmark \& Peebles \cite{Teg98} (see also \cite{Fry96}), that the
tracers and the underlying mass distribution
share the same velocity field. % and thus the same gravity field.
Using the latter and Eqs.(\ref{grper2}), (\ref{grper22}) we have
\be
\label{eq:vel}
\dot{\delta}_{\rm tr}-\dot{\delta}_{m}=0 \;.
\ee
Now since we assume linear biasing, i.e. Eq.(\ref{eq:1}),
we obtain:
\be
\label{eq:teg1}
\delta_{m}\frac{db}{dt}+
(b-1)\frac{d\delta_{m}}{dt}=0  \Rightarrow \frac{d(y\delta_{m})}{dt}=0
\ee
where $y=b-1$ and $\delta_{m}\propto D$.
An integration of Eq.(\ref{eq:teg1}) provides:
\be
\label{eq:teg}
b(z,k)=1+y(z,k)=1+\frac{b_{0}-1}{D(z,k)}
\ee
where $b_{0}$ is the bias factor at the present time.

This model is known to suffer from the so-called {\em unbiased} and
the {\em low redshift} problems, by which 
the bias of mass tracers which either obey $b_{0}<1$
or are located at relative large redshifts,
$z>0.5$, cannot be modeled by Eq.(\ref{eq:teg}) 
%the assumption of Tegmark \& Peebles \cite{Teg98} namely equal
%velocities is problematic.

\underline{\em Case 2: Tracers and Mass share same acceleration}
\underline{\em field:}
Now we consider that both the tracers and the underlying mass distribution
share the same gravitational field but different velocity fields
\cite{BasIni}.
Inserting Eq.(\ref{eq:1}) into Eq.(\ref{odedelta3}) and using
simultaneously Eq.(\ref{eq:12}),
we obtain a second order differential equation which describes the
evolution of the linear bias factor, $b$, between the background
matter and the mass-tracer fluctuation field:
\be
\label{eq:hdif}
\ddot{y}\delta_{m} + 2(\dot{\delta_{m}} +
H \delta_{m}) \dot{y} + 4 \pi G_{\rm eff}
\rho_{m} \delta_{m} y =0 \;,
\ee
where $y=b-1$. 
%Below, we will prove
%that the above expression is valid for any cosmological
%model
\footnote{The current theoretical approach
does not treat the possibility
of having interactions in the dark sector. Also
discussions beyond the linear biasing regime can be found in
\cite{Mc09} (and references therein).}
%including those of modified gravity, with $G_{\rm eff}=G_{N}Y(t)$.
Transforming equation (\ref{eq:hdif})
from $t$ to $a$, we simply derive the evolution equation
of the function $y(a,k)$ [where $y(a,k)=b(a,k)-1$] which has
some similarity with the form of eq.(\ref{odedelta1}) as expected.
Indeed, we have:
\be
\label{ydedelta1}
\frac{d^{2} y}{da^{2}}+\left[A(a)+\frac{2f(a,k)}{a}\right]\frac{dy}{da}+
B(a,k)y=0 \;.
\ee
In Basilakos \& Plionis \cite{BasIni,Bas01}, we have provided an approximate
solution of Eq.(\ref{eq:hdif}), using $f(z,k)=f_{0}(z)\sim 1$
(which is valid at relatively
large redshifts), for cosmological models
in the framework of general relativity,
which contain quintessence (or phantom) dark energy.
Here our aim is to provide an exact 
bias solution taking into account metric perturbations, namely $b(z,k)$,
for all possible dark energy cosmologies (for the case with no metric
perturbations see \cite{BasPl11}).

Inserting now $y(a,k)=g(a)/D(a,k)$ into Eq.(\ref{ydedelta1})
and using simultaneously
equation (\ref{odedelta1}) and the first
equality of equation (\ref{fzz222}), we obtain:
\be
\label{qdedelta1}
\frac{d^{2}g}{da^{2}}+A(a)\frac{dg}{da}=0
\ee
a general solution of which is
\be
\label{qdelta1}
g(a)=C_{1}+C_{2} \int \frac{da}{a^{3}E(a)}
\ee
where $C_{1}$ and $C_{2}$ are the integration constants. Utilizing now
$a=(1+z)^{-1}$, $b=y+1=(g/D)+1$ and Eq.(\ref{qdelta1}),
we finally obtain the functional form which provides our general
solution for all possible types of DE models, as:
\be
\label{eq:final}
b(z,k)=1+\frac{b_{0}-1}{D(z,k)}+C_{2} \frac{J(z)}{D(z,k)}
\ee
where
\be
\label{eq:finalal}
J(z)=\int_{0}^{z} \frac{(1+x) dx}{E(x)}\;.
\ee
An extension of the above model to include the effects
%evolution of the dark matter halos due to 
of halo merging processes, which 
introduces one further component in Eq.(\ref{eq:final}),
has been phenomenologically modeled in \cite{Bas08, BasPl11}
and it was found, using cosmological N-body simulations, 
that such effects are important only for
$z\magcir 2.5-3$. Therefore, in the light of currently 
available "growth of structure" data 
(which reach $z\sim 1$; {\em WiggleZ} \cite{Blake}), the merging
term in the bias evolution model has been neglected.
%An extension of the above model to include the 
%effects of
%merging, which introduces one further component in Eq.(\ref{eq:final}),
%has been phenomenologically modeled in \cite{Bas08, BasPl11}
%and it was found that such effects are important only for
%$z\magcir 2.5-3$.

Notice that the dependence of our bias evolution model on the
different cosmologies enters through the
different behavior of $D(z,k)$, which is affected
by $\gamma$ (see equations \ref{Dz221}, \ref{Dzz}),
and of $E(z)=H(z)/H_{0}$.
Since different halo masses result in different
values of $b_{0}$, one should expect
that the constants of integration $C_{1}=b_{0}-1$ and $C_{2}$ should be
functions of the mass of dark matter 
halos (see \cite{Bas08}), assuming that
the extragalactic mass tracers
are hosted by a DM halo of a given mass. 
It is interesting to mention here that our bias model, similarly 
to most others proposed in the literature,
relate a mass tracer, being a galaxy, an AGN or a cluster of galaxies, 
with a host dark matter halo within which 
the mass tracer forms and evolves. The models 
themselves follow the linear evolution of the host 
halo and not the internal evolution of 
the astrophysical processes of the tracer. Thus the 
assumption is that the effects 
of nonlinear gravity and hydrodynamics  
(merging, feedback mechanisms, etc.) can be ignored 
in the linear-regime (see \cite{Fry96,Teg98}).
%Note that
%that some discussion related with the latter issue
%is provided in appendix A.

\begin{figure}[ht]
\mbox{\epsfxsize=8.5cm \epsffile{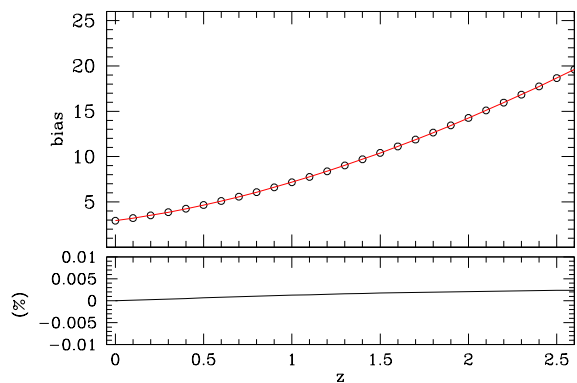}}
\caption{
The scale-dependent bias $z$-evolution ({\em upper panel}) at galaxy
cluster scales, $k\simeq 0.05 h$Mpc$^{-1}$ (open symbols), while the 
scale-independent prediction is shown as the solid line. In the lower
panel we present the fractional difference with
respect to the scale-independent bias model.
Note that we
use $\Omega_{m}=0.273$, $w_{de}(z)=-1$, $\sigma_{8,m}(0)=0.81$ and $\gamma=0.55$.}
%Finally, the constants $(b_{0}, C_{2})$ can be found in appendix A.}
\end{figure}

Comparing our solution of Eq.(\ref{eq:final})
with that of {\em Case 1}, $b(z,k)=1+(b_{0}-1)/D(z,k)$, it becomes
evident that the latter misses one of the two components of the full
solution simply because the assumption of equal velocities leads to a
first order  homogeneous differential equation for the bias
(\ref{eq:teg1}). While the assumption of equal accelerations (with
$v_{\rm tr}\ne v_{m}$) involves the full linear perturbation analysis
and thus it produces a second order homogeneous
differential equation (\ref{ydedelta1}), independent solutions of which
are $1/D(z,k)$ and $J(z)/D(z,k)$.

The further component in the bias solution, provided by the above model,
solves the the known  {\em unbiased} and the {\em low redshift}
problems, by which the Tegmark \& Peebles \cite{Teg98} ({\em Case 1})
model suffers.

\underline{\em Case 3: Tracers and Mass do not share same velocity}
\underline{\em field:}
Here we obtain a similar to {\it Case 2} bias model but using
the nomenclature of Tegrmark and Peebles \cite{Teg98}.
%bias evolution model.
We now drop the main assumption used in {\it Case 1},
that the mass-tracers and the underlying mass distribution
share the same velocity fields,
by allowing some sort of relation between the two
(matter and mass-tracers) velocity fields.
We obtain again the corresponding equation
(\ref{eq:teg1}), starting only from the continuity
equations (\ref{grper2}, \ref{grper22})
and introducing an additional time-dependent term,
$v_{f,{\rm tr}}(a)-v_{f,m}(a)$, which we
associate with the effects of the different velocity fields.
We also use the same notation, as in our original formulation,
that the tracers and the underlying mass distribution
share the same gravity (accelerations) field. Then we obtain:
\be
\dot{\delta}_{\rm tr} - \dot{\delta}_{m}=
\frac{k^{2}}{a^{2}}
\left[v_{f,{\rm tr}}(a)-v_{f,m}(a) \right]
\ee
or
\be
\frac{d(y\delta_{m})}{dt}=\frac{k^{2}}{a^{2}}
\left[v_{f,{\rm tr}}(a)-v_{f,m}(a) \right]
\ee
a general solution of which is
\be
y(a,k)\delta_{m}(a,k)=C_{1}+\int \frac{k^{2}da}{a^{3}H(a)}
\left[v_{f,{\rm t}r}(a)-v_{f,m}(a) \right] \;.
\ee
Thus the evolution of bias becomes:
\be
\label{eq:final11}
b(z,k)=1+\frac{b_{0}-1}{D(z,k)}+\frac{I(z,k)}{D(z,k)}
\ee
where
\be
\label{eq:finala11}
I(z,k)=\int_{0}^{z} \frac{(1+x)}{E(x)} \frac{k^{2}}{\delta_{m}(0,k)H_{0}}
\left[v_{f,{\rm tr}}(x)-v_{f,m}(x) \right] dx
\ee
and $b_{0}-1=C_{1}/\delta_{m}(0,k)$.
Obviously, if $v_{f,{\rm tr}}=v_{f,m}$ then
Eq.(\ref{eq:final11}) boils down to
Eq.(\ref{eq:teg}) as it should.
Although we do not have a fundamental theory to model the
time-dependent $v_{f,{\rm tr}}(a)-v_{f,m}(a)$ function,
it appears to depend on the Hubble constant $H_{0}$
as well as on the scale 
due to the fact that the 
quantity $I(z,k)$ of Eq.(\ref{eq:finala11}) has to be unit-less.
With the above in mind, we thus observe that from
Eqs.(\ref{eq:final}), (\ref{eq:finalal})
and (\ref{eq:final11}), (\ref{eq:finala11}) we obtain:
\be
v_{f,{\rm tr}}(a)-v_{f,m}(a)=\frac{C_{2}H_{0}\delta_{m}(a=1,k)}{k^{2}}
\ee
implying the following scaling of the velocity potentials
with the scale factor:
\be
v_{\rm tr}(a)-v_{m}(a)=-\frac{C_{2}H_{0}\delta_{m}(a=1,k)}{ak^{2}} \;.
\ee
In general the above difference between the velocity potentials
could have had a different form with respect to the 
scale factor, but it must always be $\propto H_0/k^2$ in order for the $I(z,k)$ 
integral to be unit-less.

Finally, in Figure 1 ({\em upper panel}) we 
compare the scale-dependent (symbols) and
scale-independent (solid curve) bias evolution 
$b(z,k)$ at galaxy cluster scales, $k\simeq 0.05 h$Mpc$^{-1}$
($r\simeq 20h^{-1}$Mpc with $M\simeq 5\times 10^{14}h^{-1}M_{\odot}$).
As it is expected, the biasing is a monotonically increasing
function of redshift. In the {\em lower panel} we show the fractional
difference between the two bias evolution results and it 
becomes quite evident that the fluctuations of the FLRW metric
do not affect the bias evolution.

\begin{figure}[ht]
\mbox{\epsfxsize=8.5cm \epsffile{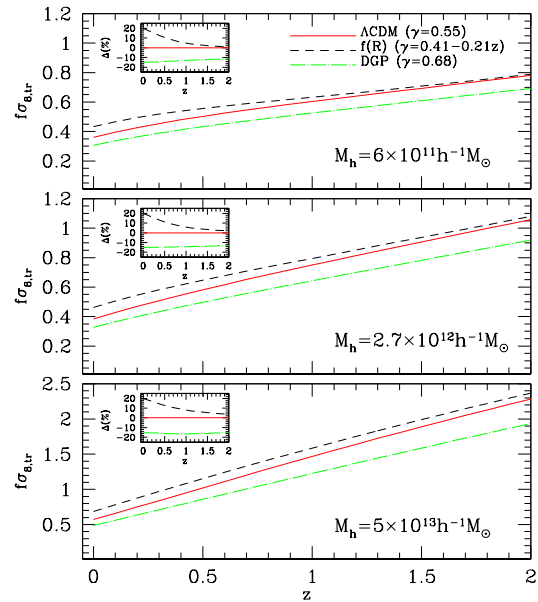}}
\caption{The predicted growth history of the Universe for different flat 
cosmological models 
({\em upper panel}: $M_{h}=6\times 10^{11}h^{-1}M_{\odot}$,
{\em middle panel}: $M_{h}=2.7\times 10^{12}h^{-1}M_{\odot}$) 
and {\em bottom panel}: $M_{h}=5\times 10^{13}h^{-1}M_{\odot}$) 
and their fractional difference with
respect to the $\Lambda$CDM model (see {\em insert panels}). 
The cosmological models shown are: $f(R)$ (dashed line) with 
$\gamma(z)=0.41-0.21z$, 
concordance $\Lambda$CDM (solid line) with 
$\gamma=0.55$ and DGP (dot-dashed) with 
$\gamma=0.68$. Note that we 
use $\Omega_{m}=0.273$, $\sigma_{8,m}(0)=0.81$. 
%\cite{komatsu11}
}
\end{figure}

\section{Testing gravity at cosmological scales}
Nowadays, the issue of testing the validity of general relativity
at cosmological scales
is considered one of the most fundamental and challenging problems on
the interface uniting Astronomy, Cosmology and Particle Physics and indeed
in the last decade there have been theoretical debates
among cosmologists regarding
the methods that researchers have to develop in order to make this
achievement.
Briefly, it is interesting to mention that measuring
the growth index could provide an efficient
way to discriminate between modified gravity models and DE models
which adhere to general relativity. Indeed it was theoretically
shown that for DE models
inside general relativity the growth index $\gamma$ is well fitted by
$\gamma_{\rm GR}\approx 6/11$ (see \cite{Linder2007},\cite{Nes08}).
Notice, that in the case of the
braneworld model of Dvali, Gabadadze \& Porrati \cite{DGP}
we have $\gamma \approx 11/16$ (see also \cite{Linder2007}), while for
the $f(R)$ gravity models we have $\gamma(z)\approx 0.41-0.21z$ for 
$\Omega_{m}=0.273$ \cite{Polarski}.

Recently, it has been proposed (see for example \cite{Vik09}) that an efficient
avenue to constrain the $\gamma$ parameter is by determining
observationally the redshift-dependent linear growth of
perturbations. Other methods have also been proposed in the
literature, such as redshift space
distortions in the galaxy power spectrum, weak lensing and the growth rate of
massive galaxy clusters (see for example \cite{Guzz08} and references therein).
%It is interesting to mention here that the above methods
%assume a linear and scale-independent bias.
Indeed, it has been shown \cite{SP09} that a good test to discriminate
among ``classical'' DE models and modified gravity models is to
compare the expected combination parameter of the growth rate of structure,
$f(z)$, and the redshift-dependent rms fluctuations of the linear
density field, $\sigma_8(z)$, with that measured observationally (from 
large redshift surveys, like the {\em WiggleZ}; \cite{Ross06,Blake} 
and references therein).

Armed, with our bias evolution model it is straightforward
to obtain theoretically a model-independent way of expressing 
the parameter combination $f(z)\sigma_{8,{\rm tr}}(z)$.
Since the metric fluctuations do not significantly affect the evolution
of bias and thus the formation of large scale structures at the scales
of interest, we utilize: $\xi_{k}(z,k)\equiv 0$.
Therefore, using the solution of the bias evolution
eq.(\ref{bias2}) and 
eq.(\ref{fzz222}) we find that the growth history of the Universe is given by:
\be
\label{eq:form}
f(z) \sigma_{8,{\rm tr}(z)}(M_{h},z)=\Omega^{\gamma}_{m}(z) b(M_{h},z) \sigma_{8,m}(z)
\ee
where $\sigma_{8,m}(z)=\sigma_{8,m}(0)D(z)$ and $b(M_{h},z)$ is given by 
eq(\ref{eq:final}). 
%with $\sigma_{8,m}(0)$ being the 
%rms mass fluctuation on $R_{8}=8 h^{-1}$ Mpc scales at redshift $z=0$. 
Our proposed gravity test consists in comparing the above expectation
with observationally determined values of 
$f(z)\sigma_{8,{\rm tr}}(z)$, 
estimated by using existing or future redshift catalogs of extragalactic mass tracers
(large red galaxies, optical or X-ray QSO, clusters of galaxies, etc).
Note, that such data are already 
available in the literature for the case of bright emission-line
galaxies \cite{Blake}.

Evidently, the essential cosmological parameters that enter in the
theoretical expectation of Eq.(\ref{eq:form}) are:
$\Omega_m, w_{de}(z), \sigma_{8,m}(0)$ and $\gamma$.
Note however that:
\begin{itemize}
\item there is only a weak dependence of $\gamma$ on
$w(z)$, as has been found in Linder \& Cahn \cite{Linder2007}, which 
implies that
one can separate the background expansion history, $E(z)=H(z)/H_0$,
constrained by a large body of cosmological data (SNIa, BAO, CMB), 
from the fluctuation growth history, given by $\gamma$, and
\item the value of $\sigma_{8,m}(0)$ remains relatively constant
  ($\sigma_{8,m}(0) \in [0.77, 0.81]$) for a range of dark energy
  equations of state ($\Lambda$CDM, quintessence, CPL), 
  as shown by a recent analysis of SDSS Large Red Galaxies \cite{Shan}.
\end{itemize}
In particular, the aim of our proposed method 
is to constrain, for a given expansion
history, the value of $\gamma$, and test whether there are deviations
from the GR expectations. In order to visualize the redshift
and $\gamma$ dependence of the $f \sigma_{8,{\rm tr}}$, we
compare in Figure 2 for different masses of dark 
matter halos, three flat cosmological 
models, in which we impose $\Omega_{m}=0.273$ and 
$\sigma_{8,m}(0)=0.81$. In particular, we consider the following cases: 

\noindent
(a) the $f(R)$ model  
with $\gamma(z)=0.41-0.21z$ (dashed line), 

\noindent
(b) the concordance 
$\Lambda$CDM ($\gamma=0.55$, solid line), and 

\noindent
(c) the DGP model with $\gamma=0.68$ (dot-dashed line). 

The inset panels of Figure 2 show the relative difference of
the $f(R)$ or DGP model $f \sigma_8$ parameter combination with
respect to that of the $\Lambda$CDM. 
Interestingly, the $f(R)$ models 
show the largest deviations ($\ge 10\%$) at the lower redshift end
($z\le 0.5$), while the DGP model shows large deviations also at the
higher redshift end.
It should be mentioned that qualitatively and quantitatively
the relative differences among the models are quite similar, independently of
the DM halo mass used, a fact which indicates that any extragalactic mass
tracer can be used with the same efficiency to perform this
cosmological test.

We will elaborate on the details of our proposed
method in a forthcoming paper, but it is important to realize 
that the parameters ($C_2, b_0)$ of our bias model 
depend: (a) on the characteristic DM halo mass, within
which the mass tracer is embedded (see Appendix A), and (b) on the values of
$\Omega_m$ and $\sigma_{8,m}(0)$ (see for example,
\cite{papag12}).
In any case  we expect that the variation of the bias model ($b_0, C_2$) parameters, 
%given in the Appendix for the WMAP7 $\Lambda$CDM model, 
within a physically acceptable range of $\Omega_m, \sigma_{8,m}(0)$
values, should be quite small. For example, the fact  that
$b_{0} [\propto 1/\sigma_{8,m}(0)]$ with $\sigma_{8,m}(0) \in [0.77,
0.81]$ (as indicated in \cite{Shan}) implies that $b_{0}$
remains mostly unaffected as far as its 
dependence on $\sigma_{8,m}(0)$ is concerned.

\section{Conclusions}
In the current work we provide a general bias
evolution model, based on linear perturbation theory,
which takes into account also metric fluctuations.
We find that the metric fluctuations do not affect the evolution
of bias and thus the formation of large scale structures.
We argue that the evolution of the mass-tracer fluctuations,
quantified by their variance on some smoothed scale,
can be used 
to test the validity of general relativity on cosmological scales.
We would like to remind the reader that in Basilakos et al. \cite{BasPl11}
paper we have derived the evolution of bias
within the context of scale-independent bias.
The combination of the latter and current works provides a
complete investigation of the linear bias evolution issue
in cosmological studies. 
We show in our current work that the use of the combination parameter of 
the growth rate of structure and the rms fluctuations of the linear
density field could provide an efficient avenue to discriminate among 
''geometrical'' (modified gravity) dark energy
models and those that adhere to general relativity.

It is however important to spell out
clearly which are the basic
assumptions of our model, which are common also to many other bias
models in the literature:
(a) Hubble scale GR effects taken into account in the fluctuation growth
(b) the Newtonian gauge approach is employed
(c) the mass tracers and the underlying mass share the same
gravity field but different velocity fields, in contrast
to the bias model proposed by Tegmark \& Peebles \cite{Teg98} in which
they proposed the they share the same velocity field.
(d) the biasing is linear on the scales of interest, and
(e) that each DM halo is populated by one
extragalactic mass tracer, which is an assumption that enters,
at the present development of our model,
only in the comparison of our model with
observational bias data and not in the derivation of its functional form.
Finally, we assume unimportant halo merging effects 
which is quite accurate for $z<3$.

\vspace {0.4cm}

{\bf Acknowledgments.}
The authors are grateful to C. Ragone-Figueroa for providing us
with the N-body simulations.
SB wishes to thank the Dept. ECM of the
University of Barcelona for the hospitality, and the financial support from the
Spanish Ministry of Education, within the program of Estancias de
Profesores e Investigadores Extranjeros en Centros Espanoles (SAB2010-0118).
JBD acknowledges support from the U.S. Department of Energy and Arizona
State University. SD wishes to thank Vanderbilt University for
hospitality.
LP also acknowledges support through
a research Project which is co-financed by the European Union - European
Social Fund (ESF) and Greek national funds through the Operational
Program "Education and Lifelong Learning" of the National Strategic
Reference Framework (NSRF) - Research Funding Program: THALIS - Investing
in the society of knowledge through the European Social Fund.
MP acknowledges funding by Mexican CONACyT grant 2005-49878.  

\appendix
\section{Parametrizing the Bias Evolution Model using N-body 
Simulations}
Our analytical solution Eq.(\ref{eq:final}) gives a family of 
dark matter (DM) halo bias
curves with two unknown parameters ($b_{0}, C_{2}$), which depend
on the halo mass as well as on the cosmological background (see
\cite{Bas01} and the Appendix in \cite{papag12}).
One can determine the behavior 
of the linear bias factor as a function redshift and halo mass, 
evaluating these constants using, for example, N-body simulations.
To this end we use the results of the high resolution, collisionless, 
WMPA7 $\Lambda$CDM simulation of  \cite{papag12}.
Here we only present the basic
information regarding this N-body simulation.

The simulation is a random realization of the concordance $\Lambda$CDM
model \cite{komatsu11} with a volume of a 500 $h^{-1}$ Mpc cube and
$512^3$ particles. The adopted cosmological parameters are:
$\Omega_m=0.273$, $\Omega_\Lambda=1-\Omega_m$, $h=0.704$ and
$\sigma_8=0.81$, while the particle mass is $7.07 \times 10^{10}
h^{-1} M_{\odot}$
comparable to the mass of a single galaxy.
%The initial conditions were generated using the GRAFIC2
%package (Bertschinger 2001).
The simulation was performed with an updated version of the parallel
Tree-SPH code GADGET2 \cite{Springel2005}.

Furthermore, the details of the method used to identify the DM halos 
and estimate their bias, as a function of redshift, with respect to
the underlying mass distribution have been presented elsewhere (eg.,
\cite{ragplio} and references therein). We only
mention here that 
the DM halos are defined using a Friends of Friends algorithm with a linking
length $l=0.17\langle n\rangle^{-1/3}$, where $\langle n \rangle$ 
is the mean particle density and that the DM halo bias is estimated by
measuring the
ratio of the variance of the DM halo fluctuations to that of the
underlying DM in spheres of 8 $h^{-1}$ Mpc radius, while its uncertainty
is based on the bootstrap re-sampling technique.

For our present analysis we use 10 different redshift snapshots of 
the N-body simulation, spanning the range: $0\le z \le 5$, while the
DM halo catalogs are determined for five different halo mass
intervals (as in \cite{ragplio}).

In order therefore the estimate the constants $b_0$ and $C_2$ 
%as a function of of redshift and DM halo mass and 
within the context of the specific cosmological model used, we fit the
DM halo bias results of the N-body simulation to the
corresponding  theoretical bias formula (\ref{eq:final}) and find
accurate fitting formulas for both parameters.
% (within a physical range of cosmological parameters).
These are
\be 
b_{0}(M_h)=C_{\beta}\left[ 1+ \left(\frac{M_h}{10^{14} h^{-1} M_{\odot}}
\right)^{\beta}\right]
\ee
with $C_{\beta}=0.857\pm 0.021$ and $\beta=0.55 \pm 0.06$.
and  
\be 
C_{2}(M_h)=C_{\mu}\left( \frac{M_h}{10^{14} h^{-1} M_{\odot}}
\right)^{\mu}
\ee
with $C_{\mu}=1.105 \pm 0.018$ and $\mu=0.255 \pm
0.005$. In figure 3 we show the simulation based values of the $b_0$ and
$C_2$ parameters as a function of DM halo mass together
with the best fitted functions, provided in eq. (A.1) and (A.2).

\begin{figure}[ht]
\mbox{\epsfxsize=8.5cm \epsffile{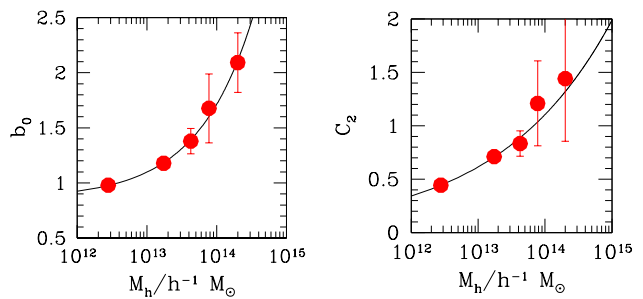}}
\caption{The values of $C_2$ and $b_0$ as a function of DM halo mass
  and the corresponding best fit curves given by eq.(A.1) and (A.2).}
\end{figure}

The cosmological dependence of these parameters is the subject of work
in progress.
%Note that using another realization of the $\Lambda$ cosmology 
%$(\Omega_{m}=0.30$, $\Omega_{\Lambda} = 0.70$, $h=0.72$ and 
%$\sigma_8=0.9$) we find that the only parameters that change 
%are $M_{*}\simeq 1.42\times 10^{15}h^{-1}M_{\odot}$ 
%and $C_{\mu} \simeq 0.51$.  

\end{document}